# Room-Temperature Nonlinear Hall Effect in Polycrystalline Magnetic Multilayers

Preet Kamal[1], Amiya Mondal[2], Kanupriya[1], Chirag Kalouni[1], Awadhesh Narayan[2] and Debangsu Roy*[1]

[1] *Indian Institute of Technology, Ropar, Department of Physics, Rupnagar 140001, Punjab, India*

[2] *Solid State and Structural Chemistry Unit, Indian Institute of Science, Bangalore,560012, India*

We report a zero-field nonlinear Hall effect in sputtered polycrystalline Ta/Pt/[Pt/Co/Fe/Ir]$_5$/Pt multilayers from 2 to 300 K. Angle-resolved transport reveals macroscopic in-plane anisotropy and a second-order response. The transverse voltage scales quadratically with current and remains unchanged across magnetization reversal. Conductivity scaling reveals strong scattering contributions, and first-principles calculations yield a symmetry-allowed Berry curvature dipole. Our findings support a mixed intrinsic-extrinsic response and establish a scalable platform for nonlinear electronics.

The Hall effect has long served as a foundational macroscopic probe of electronic band structure [1], traditionally requiring the breaking of time-reversal symmetry ($\mathcal{T}$) via external magnetic fields or intrinsic magnetization[2]. Beyond linear order, this restriction is relaxed. The in-plane current density is

$$j_a = \sigma_{ab}E_b + \chi_{abc}E_bE_c + \mathrm{O}(E^3) \quad (1)$$

where $\chi_{\mathrm{abc}}$ is the second-order nonlinear conductivity tensor responsible for both rectified dc current and a frequency-doubled (2ω) transverse response in a non-magnetic conductor at strictly zero magnetic field, the Non-linear Hall effect[NLHE] [3-5]. Sodemann and Fu showed that in the semiclassical and $\mathcal{T}$-invariant limit, the entire second-order Hall response of a 2D conductor is fixed by a single Fermi-surface quantity, the Berry curvature dipole (BCD) [3,6,7]. Notably, a second-harmonic Hall voltage by itself, however, does not establish a BCD, as intrinsic and extrinsic mechanisms can generate the same electrical harmonic and must be separated by symmetry and scaling [7,8]. The stringent symmetry constraints on the BCD largely determine the class of materials in which a BCD-driven NLHE can occur [3,9]. Systems retaining twofold or higher rotational symmetry about the film normal cannot support an in-plane BCD, irrespective of the magnitude of their local Berry curvature [10]. This symmetry suppression applies, for example, to the $C_{3v}$ Rashba semiconductor BiTeBr[11], the trigonal (111) surfaces of Bi [12] and topological insulator $Bi_2Se_3$ [13] , and the R3m ferromagnet $Fe_5GeTe_2$ [14]. A single vertical mirror can retain one BCD component, fixed normal to the mirror line, whereas in $C_1$, both in-plane components are allowed[10]. This is why pronounced NLHE is most commonly encountered in low-symmetry systems such as few-layer $WTe_2$ [15], $TaIrTe_4$ [16] and $NbIrTe_4$ [17], and in materials whose symmetry is deliberately lowered by strain, twisting, or interfacial misalignment. The essential point is that low symmetry determines which second-order channels are permitted, but it does not, by itself, determine whether the measured response is intrinsic or scattering-driven [10,11,15]. Furthermore, the NLHE is often temperature-limited; many reported observations have therefore been performed at low temperature, while room-temperature NLHE remains much less common [12,15,17]. In every such case, the materials are typically exfoliated flakes or epitaxial films with devices aligned along specific crystallographic directions [11,16,17]. Such approaches are difficult to translate directly to wafer-scale, back-end-of-line fabrication required for applications such as zero-bias rectifiers, terahertz detectors, frequency doublers, and radio-frequency energy harvesters [11,16,18].

A notable exception to the predominantly single-crystalline and epitaxial NLHE platforms is sputtered polycrystalline Bi, in which a room-temperature nonlinear Hall response was observed in (111)-textured films [12]. This result establishes an important precedent: polycrystalline averaging does not preclude a sizeable NLHE, and disorder together with surface or interface inversion breaking can sustain nonlinear Hall transport even at room temperature [12,18].

This observation is particularly relevant to sputtered metallic heterostructures, where grain boundaries and interfacial intermixing provide strong disorder [19], while inversion-asymmetric interfaces can host substantial spin–orbit coupling [20]. Disorder can generate side-jump and skew-scattering contributions to the NLHE [7], whereas Rashba-type spin splitting at interfaces can create a strongly varying Berry curvature near avoided band crossings, thereby enhancing intrinsic as well as scattering-mediated nonlinear responses [21]. Taken together, these strands suggest a concrete design rule. A conductor that simultaneously provides (i) strong interfacial Rashba spin–orbit coupling [20] with $\varepsilon_R \gg k_B T$, (ii) substantial static disorder to activate the intrinsic and extrinsic channels identified above [7,19], and (iii) a macroscopic in-plane point group reduced to a single mirror line or $C_1$ symmetry [18], should exhibit a room-temperature NLHE without recourse to exfoliation or epitaxy. Heavy-metal/ferromagnet multilayers satisfy (i) and (ii) natively, whereas the requirement (iii) is non-trivial [20].

Motivated by these predictions, we deposited thin film of *Ta(3)/Pt(5)/[Pt(1)/Co(0.5)/Fe(0.7)/ Ir(1)]$_5$/Pt(2)* (thicknesses in nm), as shown in inset of [Fig.1(a)], by off-axis dc magnetron sputtering at room temperature onto thermally oxidized Si/$SiO_2$ substrates at a base pressure of 2 × $10^{-8}$ Torr and an Ar working pressure of $4.1 \times 10^{-3}$ Torr. The [Pt/Co/Fe/Ir] (PCFI) heterostructure contains chemically and structurally inequivalent interfaces in each repeat unit Pt/Co, Co/Fe and Fe/Ir bringing together heavy elements with strong spin-orbit coupling (Pt and Ir) and ferromagnetic constituents with sizable exchange splitting (Co and Fe). Such interfaces provide a natural platform for strong interfacial spin-orbit effects, including the Dzyaloshinskii-Moriya interaction and perpendicular magnetic anisotropy, that are commonly associated with sizable interfacial Rashba coupling [22]. In addition, the asymmetric Ta/Pt seed and buffer layers beneath the PCFI core and the Pt cap above it break the global inversion symmetry and the horizontal mirror symmetry of the complete heterostructure [23]. Structural characterization by $\theta$-$2\theta$ x-ray diffraction shown in [Fig. 1(a)]. The pattern shows reflections indexed to twelve distinct planes, elucidating the polycrystalline nature of the sample. The dominant peak at 2θ ≈ 38° combines the overlapping Pt (111) and Ta (110) reflections.

Devices were fabricated by photolithography followed by etching in two complementary geometries. The first is a circular disc of diameter 40 μm with 16 radial electrodes at 22.5° intervals, that allows the drive current to be rotated continuously through 360° within a single device, [Fig.1(b)]. The second geometry is a conventional Hall bar [Fig. 1(c)] with dimensions of $360 \times 60\ \mu m^2$, patterned along the axis identified from the disc measurement and used for the quantitative characterization presented in the following sections.

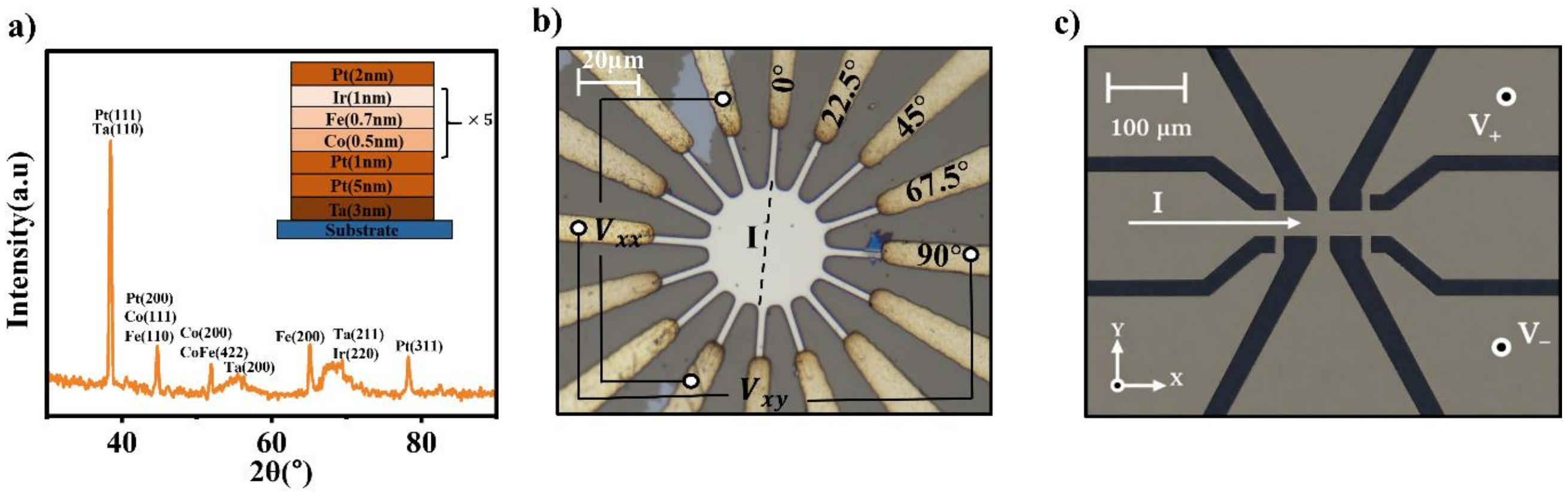

**Fig.1**: *Sample structure and device geometry. (a) XRD pattern of the multilayer stack; inset: schematic of the Ta(3)/Pt(5)/[Pt(1)/Co(0.5)/Fe(0.7)/ Ir(1)]$_5$/Pt(2) stack, Optical image of the (b) multi-terminal circular device, (c) Hall-bar contact geometry.*

We begin with the circular device [Fig. 1(b)], because it addresses the symmetry question that is specific to a polycrystalline film. An ideal polycrystalline grain is azimuthally isotropic in the film plane with no preferred in-plane direction. The first task is thus to determine experimentally whether this rotational averaging remains intact in the present multilayer before assigning a microscopic origin to the nonlinear response.

For a nominal drive direction $\theta$ in circular disc, an AC current is applied with Keithley 6221 current source between diametrically opposed contacts at $\theta$ and $\theta + 180°$, and the transverse voltage is measured with Standford Research SR830 lock-in amplifier across the orthogonal diametric pair at $\theta + 90°$ and $\theta + 270°$. Rotating the four-terminal configuration therefore samples the in-plane response within the same physical device. The geometry also provides a useful parity constraint. Advancing the complete configuration by 180° reverses the source and drain and simultaneously exchanges the transverse voltage probes. Consequently, the ideal rotating configuration obeys

$$V_{xy}^{1\omega}(\theta + \pi) = V_{xy}^{1\omega}(\theta), \qquad V_{xy}^{2\omega}(\theta + \pi) = -V_{xy}^{2\omega}(\theta) \tag{2}$$

(See *supplementary note S1 for detailed angular analysis*).

We first establish the linear transport anisotropy of the circular device. $V_{xy}^{1\omega}$ is linear in current over 1-25 $\mu$A, confirming an Ohmic contact (*see supplementary note S2*). The angular response at zero magnetic field is $\pi$-periodic and exhibits four zero crossings over a complete rotation, defining two approximately orthogonal preferred transport directions [Fig. 2(a)]. We define $\theta$ as the nominal current direction relative to a fixed in-plane device reference and $\theta_0$ as the orientation of a principal transport axis relative to the same reference. The transverse term thus can be expressed as

$$R_{xy}^{1\omega}(u) = R_H + \frac{\rho_2 - \rho_1}{2}\sin 2u, \qquad u = \theta - \theta_0 \tag{3}$$

where $R_H$ denotes an angle-independent Hall contribution. This angular form follows from the symmetric rank-two resistivity tensor; unequal resistivities along its principal axes produce a transverse electric field when the current is applied away from either axis. Such a response is permitted even in a material with point group $C_1$. The observed twofold modulation therefore establishes macroscopic transport anisotropy without assigning crystallographic $C_{2z}$ symmetry. For isotropic in-plane transport, the principal resistivities are equal, and this anisotropic contribution vanishes. The red curve in [Fig. 2(a)] is the fit of the Eq. (3) with $R_H = 0$ and the amplitude $(\rho_2 - \rho_1)/2$ as the only free parameter with goodness of fit $R^2 = 0.86$. Because the finite-contact disc has a non-uniform current distribution, the detailed lineshape in [Fig.2(a)] is not expected to follow Eq. (3) quantitatively. Accordingly, the angular modulation and the approximately orthogonal transport directions provide the principal evidence that polycrystalline averaging has not restored in-plane isotropy.

We next examine the second-harmonic transverse voltage, $V_{xy}^{2\omega}$, at zero magnetic field. Representative current sweeps reveal a pronounced quadratic dependence, $V_{xy}^{2\omega} \propto I^2$, establishing a finite second-order transverse voltage [Fig. 2(b)]. The inset resolves the smaller signals at selected weak-response orientations. This second-order response imposes a symmetry constraint distinct from the twofold modulation of the linear response. For a local constitutive relation,

$$j_a^{(2)} = \chi_{abc} E_b E_c$$

an ordinary $C_{2z}$ rotation reverses both the in-plane electric field and the second-order current, whereas $E_b E_c$ remains unchanged. Invariance under this operation, therefore, requires every purely in-plane component of $\chi_{abc}$ to vanish. A local, purely in-plane interpretation of the observed second-order response consequently requires an effective symmetry that excludes ordinary $C_{2z}$.

We parameterize the nonlinear response using the unrestricted $C_1$ tensor since the transport measurements do not independently establish a surviving in-plane mirror. For a spatially uniform field $\mathbf{E} = E(\cos\varphi, \sin\varphi)$, where $\varphi$ is measured from the same fixed reference axis, the transverse projection of the general second-order current takes the form

$$\frac{j_\perp^{(2)}}{E^2} = A_1\cos\theta + B_1\sin\theta + A_3\cos3\theta + B_3\sin3\theta \tag{4}$$

The BCD contribution is a restricted limit of Eq. (4). In the low-frequency regime,

$$\mathbf{j}_{\mathrm{BCD}}^{(2)} = \mathcal{C}(\mathbf{D}\cdot\mathbf{E})\,\hat{\mathbf{z}}\times\mathbf{E},\ \text{such that}$$

$$j_{\perp,\mathrm{BCD}}^{(2)} = \mathcal{C}E^2\left[D_x\cos\theta + D_y\sin\theta\right] = \mathcal{C}E^2 D\cos(\theta-\theta_D) \tag{5}$$

where $D = |\mathbf{D}|$, $\varphi_D$ specifies the BCD orientation, and $\mathcal{C}$ contains the transport prefactor. A pure BCD response therefore contains only the first angular harmonic and has two nodes separated by $180°$. A general $C_1$ second-order response need not be a pure sinusoid and can contain additional odd angular structure.

The angular voltage scan in Fig. 2(c) exhibits a pronounced directional response with additional structure beyond a single sinusoid. We describe the measured trace phenomenologically using the first and third angular harmonics in Eq. (4), with $\theta$ as the angular variable. The resulting fit gives $R^2 = 0.73$. These fitted amplitudes characterize the device response and are not directly identified with local nonlinear conductivity components because the electric field varies throughout the disc.

The parity-allowed component contains two zero crossings whose orientations are reproduced upon reversal of the current-contact configuration (Supplementary Note S6). Its antipodal node pairing follows from the imposed relation $V_{xy}^{2\omega}(\theta+\pi) = -V_{xy}^{2\omega}(\theta)$; the experimentally informative feature is the reproducibility of the nodal orientation across the measurement configurations. Together, the linear and nonlinear angular scans demonstrate that the polycrystalline multilayer retains macroscopic in-plane transport anisotropy and a direction-dependent second-order transverse response. Motivated by these observations, we fabricated Hall-bar devices to quantitatively investigate nonlinear transport at a fixed orientation. The longitudinal current channel was patterned perpendicular to the ($\theta = 0^\circ$) reference direction of the circular device, corresponding to ($\theta = 90^\circ$). The circular-device measurements show a finite second-harmonic transverse voltage at this orientation, providing an experimental basis for investigating the nonlinear response along the Hall-bar axis. The two geometries therefore serve complementary purposes: the circular device resolves the angular dependence within a single structure, while the Hall bar provides a well-defined longitudinal direction and an approximately uniform current density in its central channel. This geometry enables the measured voltages to be converted into electric fields and the nonlinear transverse susceptibility to be quantified. We next examine its current, temperature, frequency, and magnetic-field dependences, followed by conductivity-scaling analysis and comparison with first-principles calculations to assess the microscopic contributions.

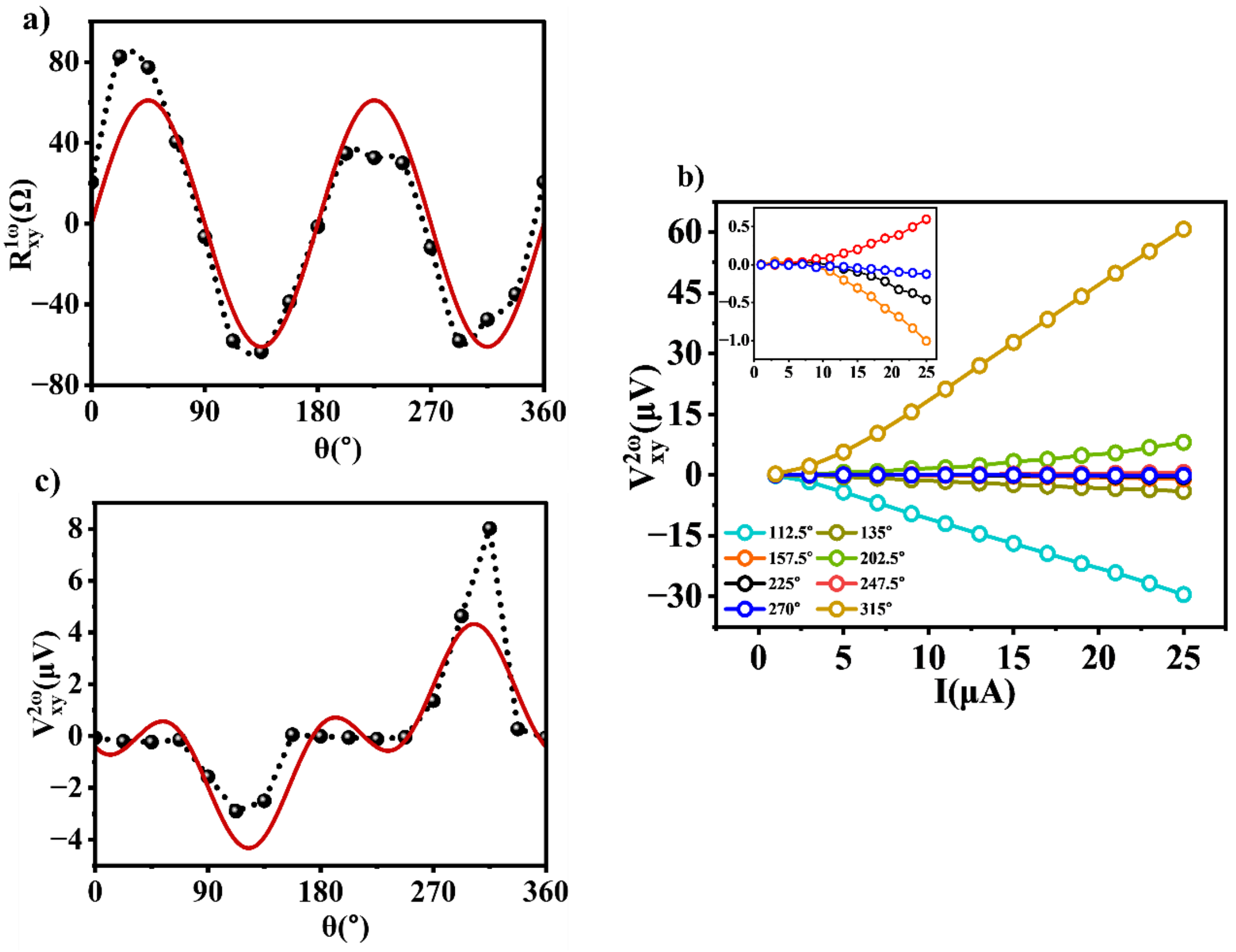

***Fig. 2**: (a) $R_{xy}^{1w}$ of the circular disk vs. $\theta$ at $5\mu A$, (b) $V_{xy}^{2\omega}$vs I at different angles,(c) $V_{xy}^{2\omega}$ vs. $\theta$ at $5\mu A$. All measurements were performed at 300 K. The connecting symbols along with dashed lines in (a) and (c) represents the experimental data. The solid red lines are the fits to this experimental data.*

In the hall bar, both $V_{xx}^{1\omega}$and $V_{xy}^{1\omega}$ vary linearly with current up to 300 K (*supplementary section S8*). $V_{xx}^{1\omega}$ exceeds $V_{xy}^{1\omega}$ by more than two orders of magnitude throughout the measured current range, confirming Ohmic transport and a well-isolated transverse voltage channel. The longitudinal slope increases systematically with temperature, consistent with metallic conduction in the multilayer.

Figure 3(a) shows $V_{xy}^{2\omega}$as a function of current at 300 K for drive frequencies between 79.571 Hz and 577.571 Hz. At every frequency, the signal follows the same parabolic dependence on current, and the traces overlap within experimental resolution. The same frequency independence is obtained at 2 K and 100 K (*supplementary section S9*). In addition, a rectified DC transverse voltage was measured using a 2182A Keithley nanovoltmeter without lock-in detection follows the same sign and overall $I^2$ trend as $V_{xy}^{2\omega}$ over the same current range, as expected for a second-order response (*supplementary section S9*). This simultaneously establishes the second-order character of the response and rules out the occurrence of the parasitic capacitive pickup or frequency-dependent electrical cross-talk, in accord with the standard controls used in $WTe_2$ [5], $NbIrTe_4$ [17] and BiTeBr [11] nonlinear-transport experiments.

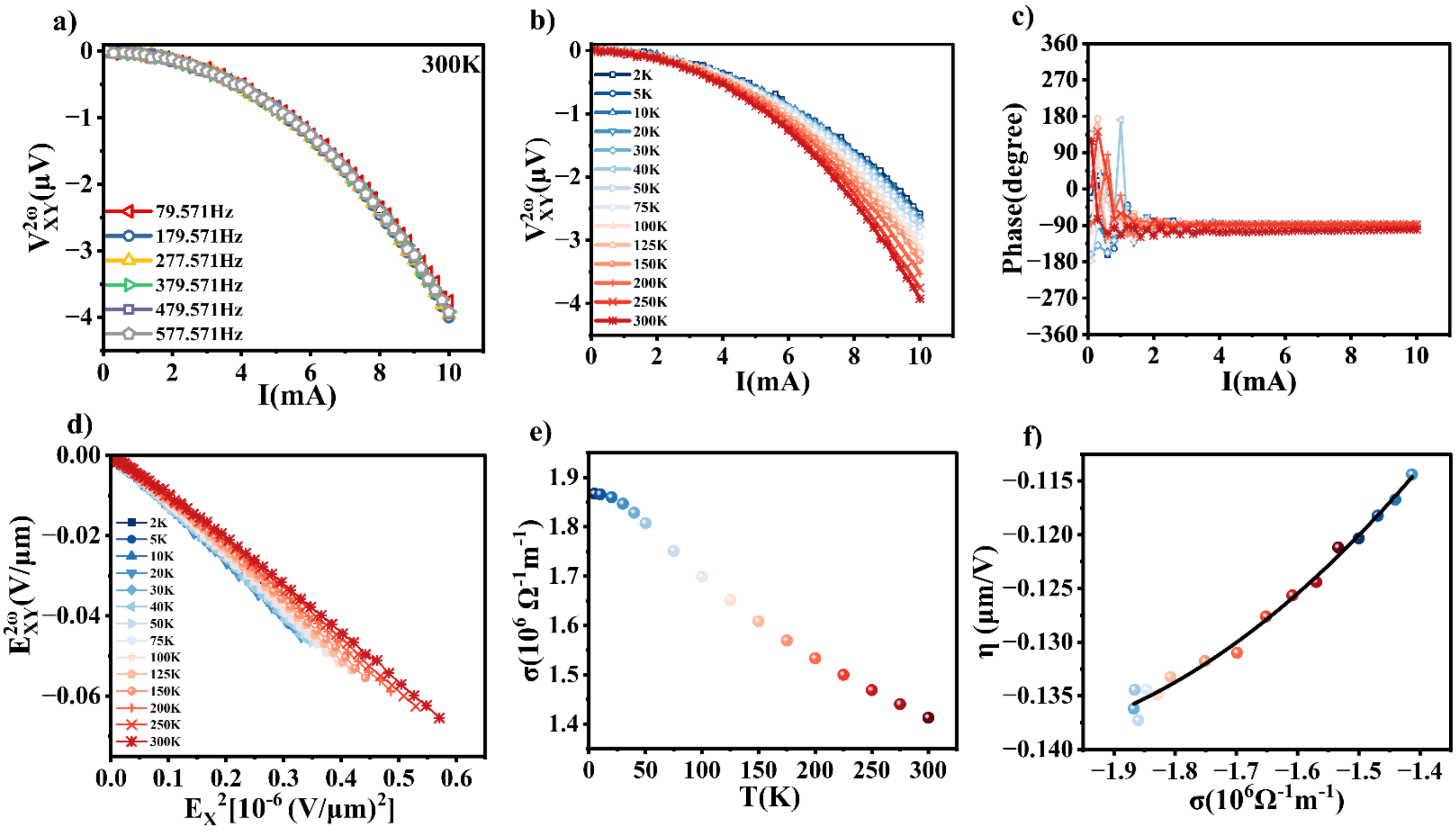

***Fig.3***: *$V_{xy}^{2w}$ vs I at* (a) *different frequencies, (b) different temperature and (c) phase of the $V_{xy}^{2w}$; each measured from 2K-300 K, (d) $E_{xy}^{2w}$ vs. $E_x^2$, (e) longitudinal conductivity σ as a function of temperature T; (c) Scaling of η as a function of longitudinal conductivity σ at a fixed driving current of I = 10 mA with fitting represented by the black solid line.*

The nonlinear transverse response persists across the complete 2 K-300 K interval as shown in [Fig. 3(b)], while its magnitude at I = 10 mA increases from 2.58 μV at 2 K to approximately 3.92 μV at 300 K. Fig. 3(c) provides an independent phase consistency check. For $I(t) = I_0 sin\omega t$, a quadratic response contains $I_0{}^2 sin^2\omega t = (I_0{}^2/2)[1 + sin(2\omega t - \pi/2)]$; under our lock-in reference convention the 2ω component is therefore expected near -90°(*supplementary section S12*) . Once the current exceeds approximately 1 mA and the signal rises above the noise floor of lock in amplifier (of order of the nV), the measured phase locks close to -90° and remains nearly temperature independent over the measured range [Fig. 3(c)]. We use this phase only as a consistency check on an instantaneous quadratic response, not as a unique fingerprint of a particular microscopic mechanism.

Next, we test whether the second-harmonic transverse response in PCFI superlattice that is magnetic in nature is tied to its perpendicular magnetic state. $V_{xy}^{2\omega}$ was measured under out-of-plane fields from -2T to +2 T at 2 K, 100 K and 300 K, spanning both the reversal region and fields well beyond perpendicular saturation. The traces recorded at -2 T, -0.13 T, -0.07 T, 0 T, +0.07 T, +0.13 T and +2 T collapse onto one another (*supplementary section S10*). Neither the magnetization reversal nor saturation in the opposite direction produces a detectable change in $V_{xy}^{2\omega}$. This excludes a dominant magnetization-odd contribution and strongly constrains anomalous Hall derived or anomalous Nernst effect backgrounds tied to magnetisation, analogous to the magnetization-reversal test used in $Fe_5GeTe_2$ [31].

Joule heating is examined separately through the transverse third harmonic. $V_{xy}^{3\omega}$ remains close to the noise floor at low temperature and reaches only about 0.25 μV at 300 K and 10 mA, more than an order of magnitude below $|V_{xy}^{2\omega}|$ at the same operating point (*supplementary section S11*). The small 3ω response shows that resistance modulation caused by periodic self-heating is too weak to account for the observed transverse 2ω voltage. Since a direct thermoelectric voltage at 2ω need not generate an equally large 3ω signal, this control is interpreted together with the frequency, phase and magnetization tests rather than in isolation. Their combined outcome supports an electronic second-order transverse transport response.

To separate the microscopic contributions to the transverse response itself, we turn to conductivity scaling. Converting the measured voltages to electric fields, $E_x = {V_{xx}^{1w}}/{L}$ and $E_{xy}^{2\omega} = {V_{xy}^{2\omega}}/{W}$, $L$ and $W$ are the length and width of the Hall bar, Fig. 3(d) shows that $E_{xy}^{2\omega}$ is linear in $(E_x)^2$ at every temperature. We therefore define the non-linear transverse susceptibility $\eta(T) \equiv E_{xy}^{2\omega}/(E_x)^2$. The magnitude of $|\eta|$ increases systematically with temperature shown in [Fig.3(f)], whereas the longitudinal conductivity magnitude decreases from $1.86\times10^6\ \Omega^{-1}m^{-1}$ at low temperature to $1.41\times10^6\ \Omega^{-1}m^{-1}$ at 300 K [Fig. 3(e)]. This continuous variation of the scattering environment provides the control parameter required for a scaling analysis. Following the established non-linear Hall scaling framework [6,7,32], the measured susceptibility is parameterized over the conductivity range as

$$\eta\,(\sigma_{\mathrm{xx}}) = a + b\sigma_{\mathrm{xx}} + c{\sigma_{\mathrm{xx}}}^2 \qquad (6)$$

As shown in Fig. 3(f), the data are well described by this form with goodness of fit $R^2$=0.99. The fit gives $a = (6.48 \pm 3.01) \times 10^{-8}\, m\, V^{-1}$, $b = (1.87 \pm 0.367) \times 10^{-13}\, m^2\, V^{-1}\, S^{-1}$ and $c = (4.31 \pm 1.107) \times 10^{-20}\, m^3\, V^{-1}\, S^{-2}$. The physically meaningful comparison is between the terms $a$, $b\sigma_{\mathrm{xx}}$ and $c{\sigma_{\mathrm{xx}}}^2$ over the measured conductivity interval. Using the signed conductivity convention of Fig. 3(f), $|b\sigma_{xx}|\approx(2.64\text{-}3.49)\times10^{-7}$ m $V^{-1}$ is the largest individual term, while $a=0.648\times10^{-7}$ m $V^{-1}$ and $c\sigma_{xx}^2\approx(0.86\text{-}1.50)\times10^{-7}$ m $V^{-1}$ partially compensate it.

Generally, the conductivity-independent contribution contains the intrinsic BCD contribution together with side-jump terms, whereas conductivity-dependent contribution depicts skew-scattering contributions. We therefore conclude that the transverse nonlinear response contains both a nonzero conductivity-independent term, in which an intrinsic BCD can reside, and strong conductivity-dependent scattering contributions.

Taken together, the Hall-bar data point to a mixed nonlinear-transport regime. At the same time, the circular-device symmetry analysis establishes a BCD-compatible polar response, and the nonzero conductivity-independent sector leaves room for an intrinsic contribution. We therefore do not claim BCD dominance from transport alone. Instead, the microscopic identification is completed by the first-principles calculation presented next, a finite BCD at the Fermi level provides independent evidence that a genuine intrinsic BCD contribution coexists with the strong disorder-mediated nonlinear response as expected in the sputtered PCFI multilayer.

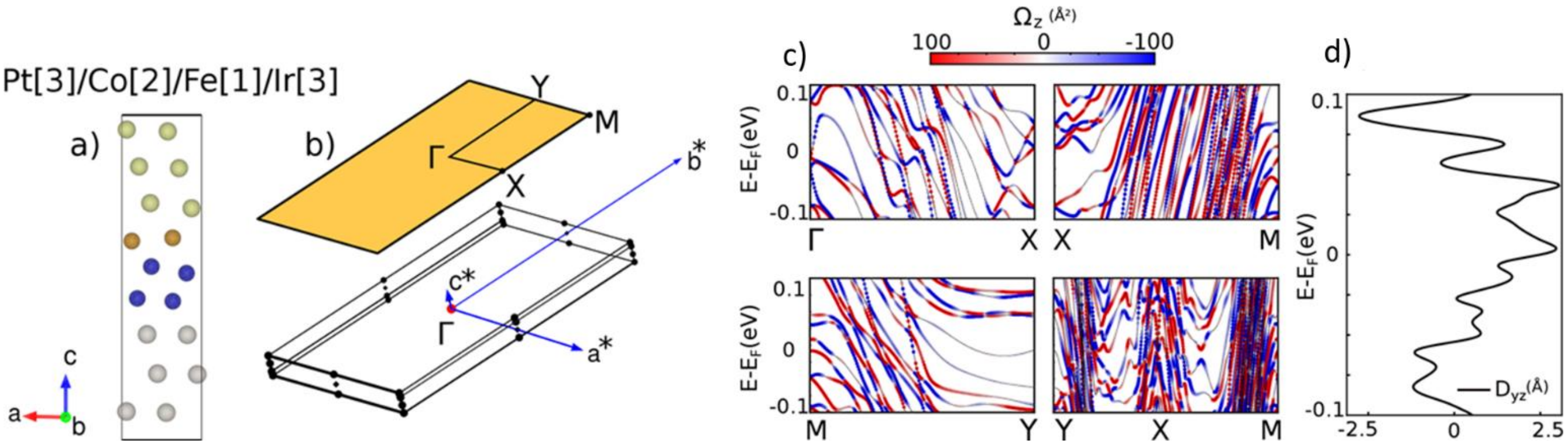

***Fig.4**: (a) Structure of the Pt/Co/Fe/Ir heterostructure showing the side view of the unit cell and (b) the corresponding Brillouin zone with the relevant high-symmetry points. The Pt/Co/Fe/Ir layers are stacked along the c axis with vaccum separating periodic images. The grey, blue, tan, and yellow spheres represent Pt, Co, Fe, and Ir atoms, respectively, (c) Electronic band structure along the high-symmetry directions, where the colour gradient represents the calculated momentum-resolved Berry curvature, (d) Calculated BCD component $D_{yz}$ as a function of energy relative to the Fermi level*

Our first-principles calculations were performed within the framework of density functional theory (DFT) using the Quantum Espresso package (*supplementary section S16*) [33]. To evaluate the Berry curvature and BCD, we constructed a tight-binding Hamiltonian based on maximally localized Wannier functions [34], using the $s, p,$ and $d$ orbitals of Pt, Co, Fe, and Ir as the initial projections. The Wannierization was performed using the wannier90 code [35]. The BCD calculations in the present study were performed using the Wannier-Berri package. (*supplementary section S16)* [36].

In this work, we consider a single PCFI multilayer stack, as our primary objective is to investigate the symmetry-allowed nonzero BCD components of the PCFI heterostructure. The multilayer stack, Ir[3]/Fe[1]/Co[2]/Pt[3], where the numbers in brackets denote the number of atomic layers, was constructed in a close-packed configuration with a (111) orientation. The Pt, Fe, and Ir layers were stacked in an fcc configuration, whereas the Co and adjacent layers were stacked in an hcp configuration. The multilayer stack was separated from its periodic images by a vacuum region of 10Å along the out-of-plane direction. The in-plane lattice constant was fixed to the calculated bulk Ir value, corresponding to a nearest-neighbor (NN) distance of 2.74Å, [Fig. 4(a-b)].

To determine the symmetry-allowed components of the BCD tensor, we performed a symmetry analysis using

$$D = \det(S)\, SDS^{-1}$$

where $\det(S)$ accounts for the pseudotensor nature of the BCD [3]. According to our DFT calculations, the pristine heterostructure possesses a single nontrivial symmetry operation, namely the mirror symmetry $M_y$. The symmetry constraints imposed by $M_y$ allow $D_{yz}$ to remain nonzero, whereas $D_{xz}$ is forced to vanish. Therefore, $D_{yz}$ is the only symmetry-allowed nonzero BCD component considered here, indicating that a BCD-induced NLHE is symmetry allowed in this heterostructure.

For our first-principles study, we considered two different types of electronic configurations. In the first configuration, SOC was included without magnetization, as shown in [Fig. 4(c-d)]. In the second configuration, magnetization was included without SOC, and the spin-up and spin-down channels were considered separately, as shown in Figs. S8 and S9, respectively (*supplementary note S17*).

The heterostructure exhibits metallic behavior, with numerous electronic bands crossing or lying close to the Fermi level, [Fig. 4(c)]. Notably a substantial Berry curvature is observed near the Fermi level, particularly in regions where the electronic bands are closely spaced. We first consider the BCD response in the presence of SOC and in the absence of magnetization. As shown in Fig. 4(d), pronounced features in $D_{yz}$ are observed within the energy window from approximately -0.1 to 0.1 eV relative to the Fermi level.

Similar finite BCD features are observed for the spin-polarized calculations without SOC, as shown in Figs. S8(b) and S9(b) of the *supplementary material*. Although the detailed energy dependence of the BCD differs between the two spin channels and the SOC case, the calculated responses remain within a comparable range in magnitude. Our results demonstrate the presence of a finite, symmetry-allowed $D_{yz}$ component in the PCFI heterostructure and therefore establish the possibility of a BCD-induced non-linear Hall response in this system.

In summary, we demonstrate a zero-field nonlinear Hall effect from 2 to 300 K in sputtered polycrystalline *Ta(3)/Pt(5)/[Pt(1)/Co(0.5)/Fe(0.7)/ Ir(1)]*$_5$*/Pt(2)* multilayer, to our knowledge the first in an engineered magnetic multilayer. Angular measurements reveal macroscopic in-plane transport anisotropy and a direction-dependent second-order transverse response, establishing that a directional nonlinear signal persists despite the polycrystalline structure. In Hall bars, the transverse second-harmonic voltage scales quadratically with current and exhibits no resolved change across magnetization reversal or over the investigated frequency range. Conductivity scaling reveals substantial scattering-dependent contributions alongside a conductivity-independent term, while first-principles calculations yield a finite, symmetry-allowed Berry curvature dipole at the Fermi level of the model heterostructure. Together, these findings support a mixed intrinsic-extrinsic mechanism. Our results establish engineered magnetic multilayers as a platform for room-temperature nonlinear Hall transport and connect this phenomenon to the interfacial electronic structure and disorder of conventional sputtered films. Because the constituent materials, sputtering process, and lithographic patterning used here are already standard in spintronic device fabrication, this route is directly compatible with wafer-scale, back-end-of-line integration, in contrast to the exfoliation or lattice-matched epitaxy required by every previously reported room-temperature NLHE system. Our findings demonstrate that the material family available for NLHE research extends well beyond van der Waals crystals and topological semimetals into the broad, technologically mature landscape of sputter-deposited magnetic thin films. This opens a practical path toward BCD and scattering-driven nonlinear electronic devices, zero-bias rectifiers, frequency doublers, and radiofrequency energy harvesters, built from conventional magnetic multilayer stacks, and suggests a broader design space in which the growth geometry, interface composition, and layer sequence of the superlattice can be used to tune the direction and magnitude of the NLHE directly.

**Acknowledgements**

Preet gratefully acknowledges the financial assistance from UGC. DR acknowledges the financial support from the Department of Atomic Energy (DAE) under Project No. 58/20/10/2020-BRNS/37125 and the Science and Engineering Research Board (SERB) under Project No. CRG/2020/005306. AM is supported by a fellowship of the Indian Institute of Science. AN thanks DST Core Research Grant (CRG/2023/000114) for support.

**Data availability**

The data that support the findings of this study are available from the corresponding author upon reasonable request